# Nonlocal Quantum Correlations: Beyond Entanglement

Mark E. Brezinski[1,2,3]

[1] Center for Optical Coherence Tomography and Modern Physics, Department of Orthopedic Surgery, Brigham and Women's Hospital, 75 Francis Street, MRB-114, Boston, MA 02115.

[2] Harvard Medical School, 25 Shattuck Street, Boston, MA 02115.

[3] Department of Electrical Engineering and Computer Science, Massachusetts Institute of Technology, 77 Massachusetts Avenue, Cambridge, MA 02139.

Email address: mebrezin@mit.edu

**Abstract**:   The utilization and control of nonlocal quantum interactions is an area of active investigation.  This is not limited to subatomic structures but extends to the macroscopic level. Nonlocal interactions can be from either entanglement or path indistinguishability (the path integral for larger systems), with the latter being further subdivided as discussed.  These two distinct phenomenon have recently been treated often in the literature as essentially identical, which is problematic when utilizing them for practical applications.  The confusion may lie in misunderstanding the physics of the type II spontaneous parameteric downconversion source (SPDS), which is used extensively with entanglement studies.  This paper examines the distinction, and why it is important for practical applications, between quantum correlations from path indistinguishability versus entanglement.  The path indistinguishability approaches discussed, under ambient conditions, are performed with a thermal source or, a coherent source (single photon wavepacket) when local entanglements in both arm have specific characteristics.  The latter we will show has the property that it can surprisingly lead to coherence expansion rather than decoherence under the proper conditions. Nonlocal quantum correlations are a complex topic that extends beyond quantum entanglement.

**Introduction**:

*General*

The ability to establish and control non-local quantum correlations would impact fields ranging from telecommunications to biology. Quantum correlations are paired measurements (in space or in time) whose results exceed that predicted by classical statistics. Unfortunately, many examples exist in the literature that treat quantum correlations from path indistinguishability and entanglement as essentially identical, an obstruction to the field and, we will argue, possibly due in part to misunderstanding of the widely used SPDC II source (spontaneous parametric down conversion)(1). Two prominent examples, in our opinion, are a 2008 *Nature* review on entanglement and the recent study claiming entanglement between two diamonds in *Science*.(2,3) A brief review of the SPDC is added to illustrate the point. But overall this paper examines the distinction, and why it is important for practical applications, between quantum correlations from path indistinguishability versus entanglement. To illustrate the concepts, two interference experiments will be examined, a HOM interferometer/SPDS and a hypothetical experiment producing nonlocal correlations using indistinguishable paths/coherent source using an M-Z interferometer, the latter under ambient conditions. The second experiment additionally demonstrates that under certain conditions first order correlations can lead to quantum correlations, and that specific local entanglements in each arm can lead to coherence expansion rather than decoherence. It is similar to the recent *Science* experiment 'entangling' diamonds (3).

*Entanglement*

Entanglement, a type of quantum correlation, is a function of superposition and the linearity of Schrödinger's equation, but not generally path indistinguishability (4). However, demonstrating interference, on the other hand, with these entangled photons does typically require indistinguishable paths, which seems to be a source of confusion. Pure entanglement (we will not be dealing with partial entanglement) is an exact correlation/pairing of base states (no matter what arbitrary basis is chosen). The entanglement process between two entities is described by (as per Von Neumman) (5):

$$|\psi\rangle|a_r\rangle = \left(\sum_i c_i |s_i\rangle\right)|a_r\rangle \rightarrow |\Psi\rangle = \sum_i c_i |s_i\rangle|a_i\rangle \quad [1]$$

The arrow represents the unitary transform resulting in the development of entanglement between a particle and the environment (or another entity), or even entanglements with fields. Entanglements can be local or nonlocal. Local environmental entanglement, for example, has for over two decades been the mechanism ascribed to decoherence (leading to a loss of either indistinguishability or nonlocal entanglement) by pioneers such as Zurek and Zen (6,7). In equation 1, the principal (which we will consider the coherent system), is given by the wavefunction (ψ) and is expressed in terms of the basis $s_i$ while the basis for the environment is expressed by $a_r$ (prior to entanglement). With two entangled particles, the two base states $s_i$ and $a_i$ develop a constant relationship (after the arrow); this is core to entanglement. Measuring one of the entangled pair establishes the eigenvalue of both exactly from the superposition. Generally, the potential of the interaction is represented in Glauber terms of an annihilation operator product (where a superposition potential is the sum of annihilation operators) (8). The initial entanglement usually requires local interaction between entities, but can be non-local with entanglement swapping, which for reference we use the well known Brune or DCLZ studies (9,10). In the Brune studies, after the initial local entanglement, the EM field and atom are in a non-local entanglement (9). Then, when a second atom is passed through the EM field after the first, the two atoms are entangled with each other. But neither is entangled with the EM field at this point.

When base states are partially paired, it becomes more controversial how much pairing is needed for the correlation to be considered entanglement. Measures such as the concurrence are used in this case, which is addressed elsewhere (11).

*Spontaneous Parametric Downconverted Source (SPDS) and Interference*

We suggest that part of the confusion between quantum correlations from entanglement and indistinguishable paths results from a misunderstanding of SPDC, light sources used widely in entanglement studies. SPDC sources generally use a CW pumped nonlinear crystal to produce two energy entangled photon pairs (including entanglement of uncertainty)(1,12). They were initially pursed to test EPR-B, generating considerable excitement because Bell-like states were produced. Due to energy conservation, photon pairs' angular frequency and wave number are entangled. According to the standard theory of parametric down conversion, the two-photon state can be written as:

$$|\Psi\rangle = \int d\omega_p A(\omega_p) \int d\omega_1 d\omega_2 \delta(\omega_1 + \omega_2 - \omega_p) a^+(\omega_1) a^+(\omega_2) |0\rangle \quad [2]$$

where ω represents the angular frequency of the signal (1), idler (2), and pump (p) of the down conversion. The $a^\dagger$ represents the respective annihilation operators and they are in a product form, consistent with an entangled state. The delta function represents perfect frequency phase matching of the down conversion (i.e. entanglement). A(ω) is related to the wavepacket extent and is not critical to the discussion here (but is for interference). This is a type I SPDC source (photons share same polarization); note that the equation does not require or include path indistinguishability. With a type II SPDC source, the signal and the idler have orthogonal polarization states (i.e. the energy entangled photons are associated with perpendicular polarizations). The state is given by (1,12)

$$|\Psi\rangle = \int d\omega_p A(\omega_p) \int d\omega_1 d\omega_2 \delta(\omega_1 + \omega_2 - \omega_p) a_o^+(\omega_1) a_e^+(\omega_2) |0\rangle \quad [3]$$

The subscripts on the signal and idler represent different polarization states associated with the entangled energy states (o and e). Again, the energy states are entangled (and thereby the polarization states) without any use of indistinguishable paths.

Now, the HOM interferometer (Figure 1) using a SPDC II source illustrates both entanglement and path indistinguishability. In this set-up, prior to the beam splitter, the photons are both entangled by energy and polarization. After the beam splitter, indistinguishable paths are present to the two detectors. Under the correct set-up of the polarizers ($P_1$ and $P_2$) in each arm, with indistinguishable paths, Bell states can be generated which can be used to test, for example, EPRB (13). But the path indistinguishability after the beam splitter does not cause the entanglement; rather it is used to generate Bell states from the entangled states (14). If the path length mismatch is substantially greater than length of the wavepackets and/or the detector times rapid, path distinguishability is lost or reduced (the Bell states are not produced), but energy entanglement is maintained. Authors often abbreviate the wave function for these Bell states (energy entangled photons grouped by indistinguishable paths) for example, as ½(|HV⟩ + |VH⟩), eliminating the energy entanglement from the equation. This representation, as seen in the *Nature* review, can be misleading, because it drops the energy/polarization entanglement that exists without the beam splitter, as well as the wavepacket for the biphoton (basically just using the e and o from equation II and giving the impression they are being entangled by the beam splitter). In the *Nature* paper, the commonly made yet incorrect statement in their Figure 1 reads, "However, in the regions where the two cones overlap, the state of the photons will be |HV⟩+|VH⟩. It is around these points that entangled photons are generated."(2) This abbreviated representation of the state ignores the fact the photons are already entangled by energy/polarization in areas outside the overlap (Equation 3). Entanglement exists in the areas outside the overlap, just not in the form of Bell states.

*Superpositions Generating Quantum Correlations.*

Quantum correlations from path indistinguishability (and more broadly the path integral) represents a distinct mechanism from true entanglement for generating nonlocal quantum correlations (15-18). Indistinguishable paths can lead to nonlocal quantum correlations through at least two mechanisms, either from a thermal source or a coherent source where nearly identical local entanglements (under the conditions below) exist in both arms. There is no correlating of base states and there is a requirement that <u>at the time of measurement </u>(though not necessarily before), paths are indistinguishable. For correlating two particles, this means the particles pathlength difference must be less than the wave packet width, or the detector time must be relatively long (17). This is in contrast to an EPR-B like experiment, for example, where two entangled spin 1/2 particles are separated in space, and determination of the state of one establishes the state of both (correlated base states). For EPR-B, even if the second particle is measured a long time afterwards and the paths are distinct, on assessing correlations of results, a correlation exists beyond a classical statistical correlation.

So, several points of path indistinguishability are distinct from entanglement. First, the paths can be distinguishable along their course, but as long as they are indistinguishable at the time of measurement, the quantum correlations will exist. Second, paths can become distinguishable (or indistinguishable) based on detector response time or path mismatches relative to the wave packet length, distinct from entanglements. Finally, though rarely discussed, we are dealing with the path integral and not just individual well defined paths. Therefore, we are actually dealing with indistinguishability of the action of two or more potentials rather than two distinct paths (18). This emphasis on the action rather than one distinct path offers a potential robustness against decoherence (local environmental entanglements), as the action can be maintained while specific paths can be lost to local environmental entanglements.

The description of entanglement above was for a two particle system, but for discussing indistinguishable paths in this section, we will focus on single particles through the paths or first order coherence (which usually is not associated with quantum effects). This is because both of the recent diamond 'entangling' *Science* paper and single photon wavepacket interference is easier to illustrate than the biphoton wavepacket (there is more than one type) (3). With resect to two particles (biphoton wavepacket), other groups and our team have published papers discussing the physics of second order correlations (SOC) from *thermal sources* demonstrating quantum correlations through path indistinguishability (15-17). In addition to the listed references, a brief summary of thermal SOC general principals is given in the appendix. Again, a first order coherence (which of course requires path indistinguishability) thought experiment or Gedankenexperiment, will be the focus, which will demonstrate that quantum correlations not only exist for thermal sources, but can occur with a coherent source under the correct conditions. This we are unaware has been addressed in the literature.

*Hypothetical Interferometer (Gedankenexperiment)*

As per Dirac, first order coherence involves interference of a single photon wavepacket with itself (while second order coherence interferes a bi-photon wave packet) (19). Different photons do not interfere with each other (19). The experiment proposed closely resembles the recent *Science* paper 'entangling' diamonds non-locally, though we will argue that entanglement exists within each arm but the nonlocal quantum correlations between arms were due to path indistinguishability (plus certain special circumstances)(13). The schematic of the interferometer is shown in Figure 2. The source is coherent and one photon enters the interferometer at a time. The M-Z interferometer uses two beam splitters and two detectors. Each arm contains identical Raman scatterers embedded in an optically transparent medium (like a diamond), absorbing the photon (for this illustration we are ignoring the high intensity requirements for significant Raman scattering), and emits identically frequency shifted photons (Stokes photons). The absorbed energy is dispersed into the object containing the Raman scatterers in the form of phonons. The properties of the Raman scatterers are specific, but before discussing them and the theoretical interferometer described, the physics of single photon interference will be introduced, a purely quantum mechanical phenomena. Again, all first order coherence (second order interference) is a single photon wavepacket interference through path indistinguishability. A more complete discussion of coherence and indistinguishability can be found in the pioneering work of investigators such as Mandel (20) and Shih (for both single and two photon [boson] correlations), including the latter's recent book on the issue (16,17). A previous paper paper by our group and a 2011 publication by Shih also describe the quantum effects of thermal second order correlations (SOC) or biphoton wavepacket (15,16).

*Interferometer Theory*

Briefly, we begin looking at path indistinguishability for a single photon entering a beam splitter. Again, all first order coherence, no matter the intensity, is single photon wavepacket interference (as per Dirac) along indistinguishable paths (19,20). In other words a photon can only interfere with itself (or biphoton interfere with itself). A single photon passing through a beam splitter is given by:

$$|\psi\rangle = \alpha|1\rangle_1|0\rangle_2 + \beta|0\rangle_1|1\rangle_2 \quad [4]$$

Here the subscripts 1 and 2 are the two paths and the value in the ket represents occupation number. We will let the alpha and beta terms take into account beam splitter ratios. We will assume polarization and dispersion are identical at the output of the beam splitter. Each arm is reflected off of mirrors, passed through phase retarders, and combined at the second beam splitter as in Figure 2. The density operator (in its expanded form) at the second beam splitter is given by:

$$\hat{\rho} = |\alpha|^2 |1\rangle_1 |0\rangle_2 \langle 0|_2 \langle 1|_1 + |\beta|^2 |0\rangle_1 |1\rangle_2 \langle 1|_2 \langle 0|_1 + [\alpha\beta^* |1\rangle_1 |0\rangle_2 \langle 1|_2 \langle 0|_1 + h.c.] \quad [5]$$

We don't need to concern ourselves with the two paths after the beam second beam splitter for now. We are only concerned that a single photon is detected by ether detector (but we still can't distinguish photons from path 1 or 2). The first two terms, the diagonal terms, are the DC terms that reduce fringe visibility to a maximum of 50% unless they can be removed (for true entanglement, there are no DC terms and maximum visibility can reach 100%). When paths are distinguishable these are the only non-zero terms. The third and fourth terms represent indistinguishable paths and generate interference (*h.c.* is the Hermitian conjugate or adjoint). These off diagonal elements are generally complex. It should also be clear that maximum interference only occurs, based on the cross terms, within the conditions that the paths are indistinguishable (terms in brackets). So for a single photon, for example, delay times in each arm need to match within the wavepacket width. Adjusting the phase delay with the polarization controllers then alters the degree of interference, guiding it to one detector or the other. In contrast, for most entangled states, delays in measurement are not an issue except when demonstrating interference.

Now we extend equation [5] beyond one photon (increase intensity) and include interactions with elements in each arm, E (either Raman scatterers or the environment). This is a relatively common procedure for explaining basic decoherence (6,7), where the relevance of the inner produce should become apparent. The interference pattern at the detector is described by the cross terms (off-diagonal) in the density operator (it is in the expanded matrix form).

$$\hat{\rho} = \frac{1}{2}\{|\psi_1\rangle\langle\psi_1| + |\psi\rangle_2\langle\psi|_2 + |\psi_1\rangle\langle\psi_2|\langle E_2|E_1\rangle + |\psi_2\rangle\langle\psi_1|\langle E_1|E_2\rangle\}$$
*where* [6]
$$|\psi_1\rangle\langle\psi_1| = \rho_{11}, |\psi_2\rangle\langle\psi_2| = \rho_{22}, |\psi\rangle_1\langle\psi|_2 = \rho_{12}, |\psi\rangle_2\langle\psi|_1 = \rho_{21}$$

The first two terms are again DC terms and the second two represent interference terms. The wavefunction (in the bras-kets) incorporates all properties of the photons (polarization, bandwidth, photon numbers, etc.) now and not just occupation number. As can be seen from the density operator, the interference pattern is independent of whether the photons come individually or at high intensity (if one of the wavefunctions were zero, interference would still occur). So the statement still holds that all interference if first order interference is single photon interference no matter the intensity. In the density operator equation, 1 and 2 correspond to the two potential paths the photon can take (ignoring for now the path integral or action). The density operator contains an inner product (E) in the last two terms that represents the Raman scattering (or any other local environmental entanglement) which can be identical or distinct between arms. The central point is that if $E_1$ and $E_2$ are approximately equal, their presence in each arm does not effect interference, even though a frequency shift occurs.

Now we go to the interferometer set up in Figure 2, which will be used in the Gedankenexperiment. The local environmental entanglements (Raman scattering in object) will have specific properties. First, $E_1$ and $E_2$ are identical as far as path indistinguishability is concerned. Second, the local interaction is Raman scattering, emitting identical frequency shifted photons in each arm. Third, Markovian and Born approximations, used in essentially all decoherence master equations, do not hold (21).

The Born approximation is that the object-incident photon interaction is sufficiently weak and environment large such that the principal does not significantly change the object in the interferometer (21). This approximations fails for the diamond experiment and others, so is not a reasonable assumption. Obviously the coupling is strong (Raman scattering) and the object changes significantly (change in phonon frequency). The Markovian approximation, having no memory effects, means that self correlations within the object/environment decay for all practical purposes instantly into the environment (21). This does not hold for the diamond study in the phonon field and in our theoretically experiment.

Now the results of this set-up are the following. First, in each arm the Stokes phonon is entangled with the phonon field <u>in that arm</u> until measurement occurs. This is analogous to the well-known Brune et. al. and DCLZ studies (9,10). Second, the phonon fields are quantum correlated not because of non-local entanglement but because of path indistinguishablity. There is no way to know which arm the photon traveled, so both phonon fields are correlated nonlocally. This correlation is from the combination of path indistinguishability from first order correlations (a source generating entangled photons is not used) and local entanglement, rather than nonlocal entanglement. Third, rather than leading to decoherence, the local entanglements under these conditions results in <u>coherence expansion and correlation between the two distant objects</u>. Fourth, indistinguishable paths can lead to nonlocal quantum correlation/coherence that does not involve nonlocal entanglement. Fifth, quantum correlations from path indistinguishablity are likely more robust to environmental changes that those from true entanglement. Both our previous work and the diamond experiment were done under ambient conditions. The quantum correlations between the phonon fields can be confirmed by following the single photon with a probe pulse down both arms, which combined anti-stokes photons in both arms (3,10).

**Conclusion**:

The paper outlines the difference between correlations from indistinguishable paths of SOC and true entanglement. *So inseparable states or density operators (in this case nonlocal quantum correlations) can be from entanglement, path indistinguishability from a thermal source, or indistinguishable states with a coherent source where identical local entanglements exists in each arm.* But another major conclusion comes from the analysis in the example of a coherent source or first order correlations with indistinguishable paths. With the indistinguishable paths of single photon coherence, near identical nature of Raman scattering, and not meeting the Born/Markovian approximations, expansion of the coherence (the two objects become part of the principal, resulting in quantum correlations) rather than decoherence. This describes why the two phonon fields become correlated and why it does not require (or include) an explanation of true non-local entanglement between arms. The paper looks at three approaches of producing SOC with inseparable density operators, emphasizing this is not a phenomenon limited to entangled photons.

**Acknowledgements**:

National Institute of Health Grants R01 AR44812, R01 HL55686, R01 EB02638/, HL63953, R01 AR6996, and R01 EB000419, 1R21EB015851.

**Appendix**:  In the text both inseparable operators with SOC form a SPDC source and with first order coherence with indistinguishable paths/local entanglement were discussed in the text,.   Thermal SOC will be discussed here.  They will not be discussed in detail here because we and other groups have discussed it in detail (15-17).  Briefly, the quantum effects of thermal second order correlations (SOC) will be expected in terms of the second order correlation function (G).  SOC can have separable (essentially accidental coincidence measurements) or non-separable density operators (quantum correlated).  Until recently, it has been strongly argued that thermal SOC (other than entangled photons) were classical statistical correlations (separable density operators).  But the thermal SOC have nonseparable density operators, and produce quantum correlations by interfering two photon wave packets through indistinguishable paths.  In modeling the two photon probability amplitude, for simplicity, the photon is going to be destroyed at the detectors (absorption).   From the quantum

theory of photodetection, the second order correlation function is given by(8):

$$G^{(2)}(t_1, r_1; t_2, r_2) = Tr[\hat{\rho} E_1^{(-)}(t_1, r_1) E_2^{(-)}(t_2, r_2) E_2^{(+)}(t_2, r_2) E_1^{(+)}(t_1, r_1)] \quad (1A)$$

Here $Tr$ is the trace and is the density operator. The two non-Hermitian operators, and , are the positive and negative frequency components of the electric field operator. The electric field operator can be expressed more explicitly as:

$$\hat{E}^{(+)}(r,t) = (\hat{E}^{(-)}(r,t))^\dagger$$
$$= i \int_0^{+\infty} (\frac{\hbar \omega_k}{2\varepsilon_0 V})^{\frac{1}{2}} \vec{e}_k \hat{a}_k \exp[i(nk \bullet r - \omega_k t)] dk \quad (2A)$$

Here $n$ is the composite refractive index of any material passed through (including the beam splitters and polarizers), is the polarization vector, v is the volume, and is the annihilation operator[15]. We will be modeling the thermal second order correlations as an incoherent statistical mixture of two photons with equal probability of having any momentum q and q' (22). Then the density operator of the second order correlations can be written as:

$$\hat{\rho} \propto \sum_q \sum_{q'} |1_q 1_{q'}\rangle\langle 1_q 1_{q'}| \quad (3A)$$

The spatial portion of the second order correlation function can be written using equations 1A and 3A as (where $x_1$ and $x_2$ are the positions of the two detectors) (22):

$$G^{(2)}(x_1; x_2) = \sum_{q,q'} |1_q 1_{q'}\rangle E_1^{(-)}(x_1) E_2^{(-)}(x_2) E_2^{(+)}(x_2) E_1^{(+)}(x_1) \langle 1_q 1_{q'}|$$
$$= \sum_{q,q'} |\langle 0| E_2^{(+)}(x_2) E_1^{(+)}(x_1) |1_q 1_{q'}\rangle|^2 \quad (4A)$$

The electric field operator is now rewritten in terms of the annihilation operator:

$$\vec{E}_j^{(+)}(x_j) \propto \sum_q f_j(x_j; q) \hat{a}(q) \quad (5A)$$

Here $j$ = 1 or 2, represent the sample arm or reference arm. The annihilation operator corresponds to a given mode and $f(x_j; q)$ is a spatial distribution function. Now substituting the field operators into equation 4A, the second order correlation function becomes (15,22):

$$G^{(2)}(x_1; x_2) = \sum_{q,q'} |f_2(x_2; q) f_1(x_1; q') + f_2(x_2; q') f_1(x_1; q)|^2 \quad (6A)$$

The interference here is not due to the superposition of electromagnetic fields as in classical optics at a focal point in space time. It is due to a superposition of the terms on the right side of the summation. This expresses the nonclassical nonlocality of the experimental observations. Interestingly, this equation can be rewritten in terms of first order correlation functions (though it is a second order effect):

$$G^{(2)}(x_1; x_2) \propto \sum_q |f_1(x_1;q)|^2 \sum_{q'} |f_2(x_2;q')|^2 + \left| \sum_q f_1^*(x_1;q) f_2(x_2;q) \right|^2$$

$$= G_{11}^{(1)}(x_1) G_{22}^{(1)}(x_2) + \left| G_{12}^{(1)}(x_1; x_2) \right|^2$$

(7A)

The events are occurring at two independent locations even though they are expressed in terms of first order correlations. An intriguing aspect of equation 7A is that the second term, generated from thermal radiation, represents a coherent superposition. Nonlocal quantum correlations can be produced from a thermal source.

**Figures**

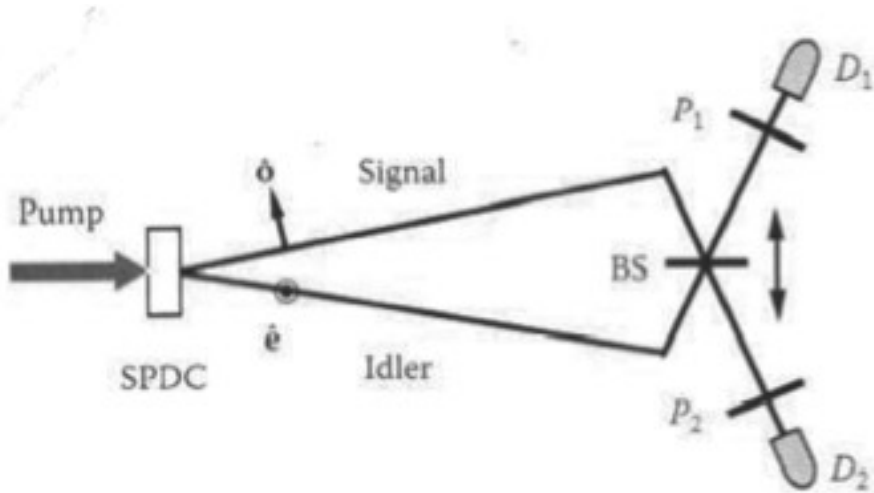

*Figure 1*: An SPDC type II source using a beam splitter to generate Bell states.

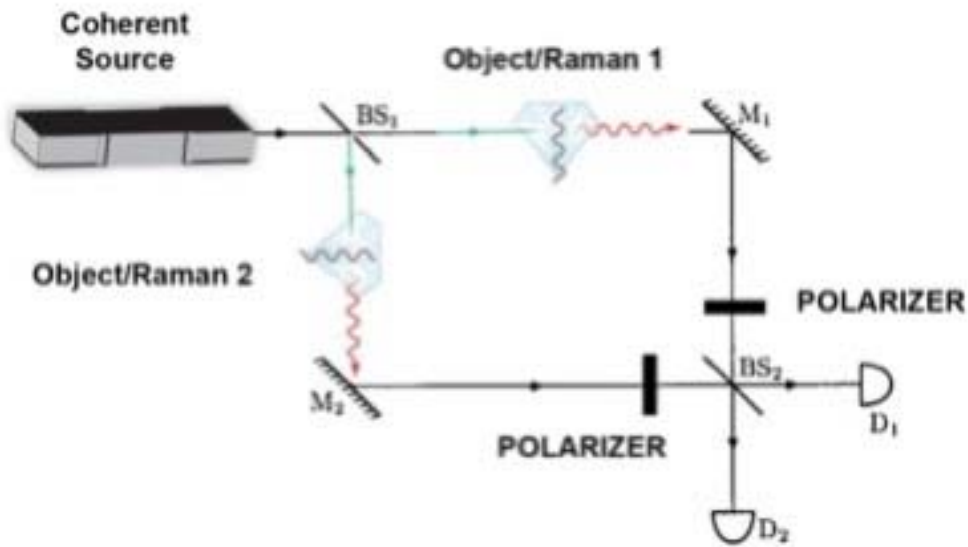

*Figure 2*: Nonlocal phonon correlations from single photon wavepacket through indistinguishable paths.